\documentclass[twocolumn,aps,prl,showpacs,raggedbottom,nobalancelastpage,amssymb,superscriptaddress]{revtex4-1}

\usepackage{graphicx}
\usepackage{amsmath}
\usepackage{amssymb}
\usepackage{bm}
\usepackage{color}
\usepackage{amsfonts}
\usepackage{appendix}
\usepackage{color}
\usepackage{dsfont}
\usepackage{soul}
\setstcolor{red}

\definecolor{lightgrey}{rgb}{0.7,0.7,0.7}

\newcommand{\ignore}[1]{\relax}

\begin{document}
\title{
Incoherent scatterer in a Luttinger liquid: a new paradigmatic limit
}
\author{Alexander Altland}
\affiliation{Institut f\"ur Theoretische Physik, Universit\"at zu K\"oln, K\"oln, 50937, Germany.}
\author{Yuval Gefen}
\affiliation{Department of Condensed Matter Physics, The Weizmann Institute of Science,  Rehovot 76100, Israel.}
\author{Bernd Rosenow}
\affiliation{Institut f\"ur Theoretische Physik, Universit\"at Leipzig, D-04103, Leipzig, Germany.}

\date{April 17, 2012}
\begin{abstract}
  We address the problem of a Luttinger liquid with a scatterer that
  allows for both coherent and incoherent scattering channels. The
  asymptotic behavior at zero temperature is governed by a new stable
  fixed point: a Goldstone mode dominates the low energy dynamics,
  leading to a universal behavior. This limit is marked by equal
  probabilities for forward and backward scattering. Notwithstanding
  this non-trivial scattering pattern, we find that the shot noise as
  well as zero cross-current correlations vanish. We thus present a
  paradigmatic picture of an impurity in the Luttinger model,
  alternative to the Kane-Fisher picture.
\end{abstract}
\pacs{}

\maketitle
The electron-electron interaction manifests itself in a particularly
pronounced way in 1D systems, inducing a strongly correlated electronic
state depicted by the Luttinger liquid (LL) models
\cite{Luttinger}. Experimental manifestations of the latter are
ubiquitous and include carbon nano-tubes, semiconductor etched edges,
polymer nanowires, quantum Hall edges and more. Early on in the
history of this field it has been realized that the presence of even a single weak
impurity in such systems gives rise to dramatic effects \cite{single_impurity}. This
has been established in the seminal work of Kane and Fisher \cite{KF}
who studied the scaling of impurity induced backscattering in the
regime of repulsive electron-electron interaction. The emerging
picture has been generalized to the chiral edges of quantum Hall
setups, and was extended to include observables such as shot noise. In
short, the Kane-Fisher (KF) picture implies that there
are two asymptotic limits of a backscattering impurity: the vanishing
impurity strength (represented by an unstable fixed point): impinging
particles are scattered forward (we refer to this as a $(a,1-a)$
splitter with $a=0$ being the backscattering probability.) The other
limit corresponds to the infinite strength impurity (a stable fixed
point): the impinging current is all backscattered, i.e. a $(1,0)$
scatterer.

\begin{figure}
\includegraphics[width=8cm]{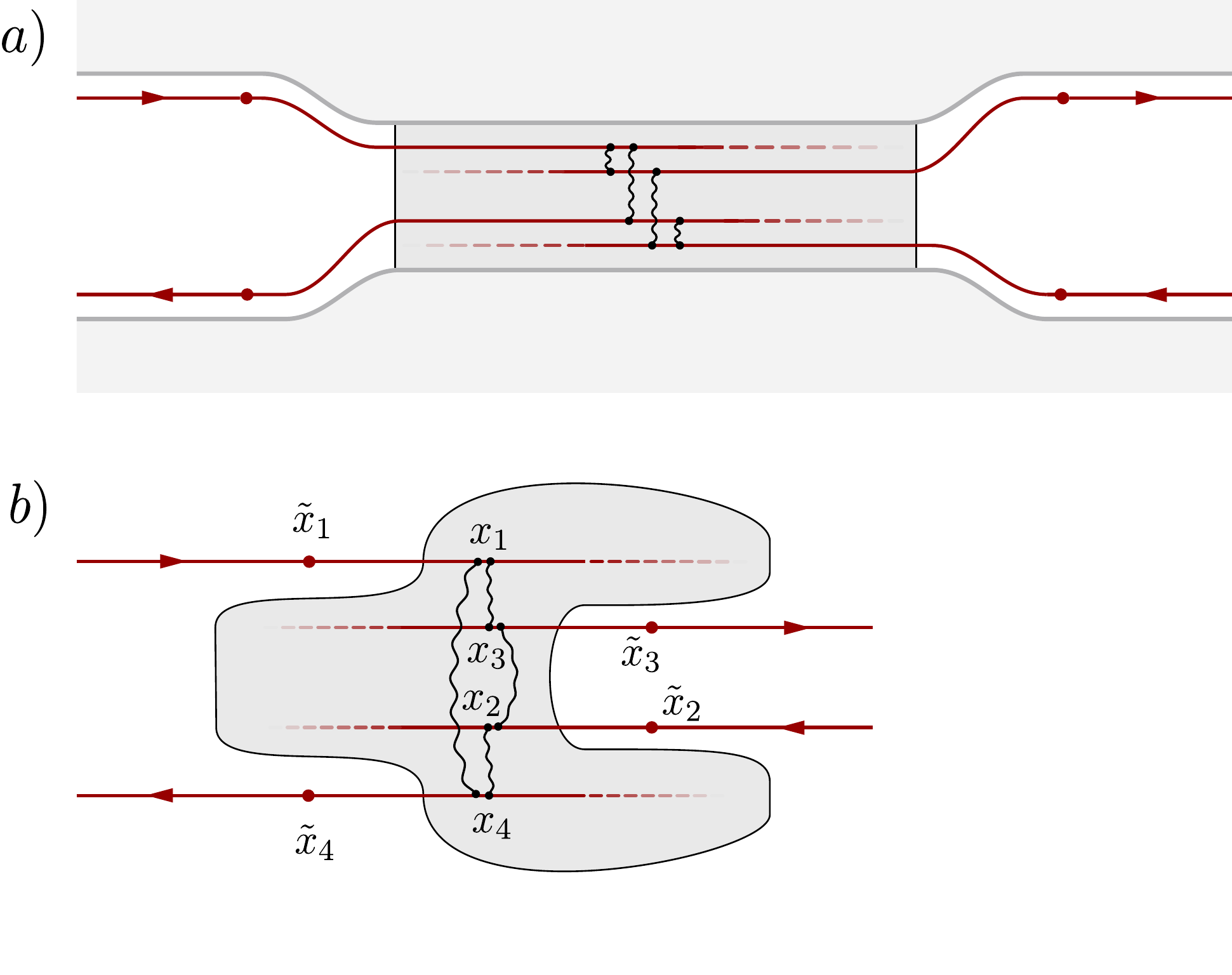}
\vspace*{-.5cm}
\caption[]{
(Color online) Quantum wire comprising two incoming ($l=1,2$) and two
outgoing $(l=3,4$) chiral
modes connecting to an extended scattering region. Wiggly lines denote
coherent scattering channels, dark shading represents a region of
capacitive charging (the quantum dot). (a) Cartoon of possible real space
structure in FQHE geometry subject to gating, (b) alternative view
adjusted to our modeling with its tunneling 'hot-spots' $x_i$ and lead
observation points $\tilde x_l$.}
\label{fig:1}
\end{figure}

The 'impurity' in the KF setup represents the paradigmatic
limit of an elastic quantum scatterer, perturbing a system of two fully
coherent left and right moving modes. The complementary limit of fully
incoherent scattering was studied by Furusaki and Matveev~\cite{Matveev}  in a model
where charge transmission was solely due to inelastic excitation
('inelastic co-tunneling') of a connecting quantum dot. In general,
however, scattering regions in quasi one-dimensional conductors may
comprise both coherent and incoherent channels of transmission and
reflection, which leads to setups intermediate between the two limits
above. For example, the action of gates on a quantum Hall bar may
effectively form compressible `quantum dots', which arguably support
both, elastic scattering channels, and inelastic scattering via bulk
gapless excitations (cf.~Fig.~\ref{fig:1}), for an experimental
demonstration see \cite{Roddaro+05}. Similarly, the
counter-propagation of edge modes along graphene pn-junctions
\cite{snake} is governed by a combination of single particle
scattering and mode interaction, which should again lead to admixtures
of coherent/incoherent transmission.

In this Letter, we explore the physics of a scattering region in which
all symmetry-allowed scattering channels between two incoming and two
outgoing chiral quantum wires are present (cf. Fig. \ref{fig:1}.)  We
will show that the low energy properties of this system differ
profoundly from those of the KF paradigm, and related
models~\cite{Nayak-multiple-lead,Egger,Chamon,Barnabe-Theriault,
  Das-Rao1,Giuliano,Bellazzini,Das-Rao2,Das-Rao3,Oreg}. Specifically,
we find that its conduction properties are governed by a stable fixed
point at which the system becomes a $({1 \over 2},{1 \over 2})$
scatterer. This limit is robust and insensitive to details concerning the
leads/scatterer coupling. In the vicinity of the fixed point, the
model shows a number of remarkable features. Most important, and
notwithstanding the fact that we deal with a non-trivial scatterer
(with a finite reflection amplitude), there is neither diagonal
  nor cross-current shot noise. These features reflect the presence
of a single gapless mode in the problem, which is protected by
symmetry and evolves in a linear, and hence noiseless manner.

Our model is depicted in Fig.~1a. Two (chiral) incoming and
two (chiral) outgoing channels are coupled to a scattering  'quantum
dot' (QD). Depending on the context, the incoming chiral channels may
represent edge modes of a fractional quantum hall bar, or the
effectively left and right moving modes~\cite{FisherGlazman} of an
interacting quantum wire.   Charge excitations arriving
in the QD may create outgoing charge, either by direct quasiparticle
scattering, or indirectly, via the creation of charge excitations on
the dot (cf.
Fig.~1b.) Our quantitative modeling, inspired by earlier work by
Furusaki and Matveev, is described by the Keldysh action
$S=S^{(0)}+S_{C}+S_{\mathrm{tun}}+S_V.
$
Here  $S^{(0)}=\sum_{l=1}^4 S_{l}^{(0)}$ represents the
quadratic bosonic action of the chiral wires,

\begin{align}
\label{eq:2}
  S_{l}^{(0)}[\phi]  ={ \hbar \over 2}\int {dq\over 2\pi} \int {d\omega\over 2\pi}\,
 \phi^T_{i,q,\omega}\,\hat G^{-1}_{q,\omega}\,
    \phi_{i,-q,-\omega},
\end{align}
where the fields $\phi\equiv(\phi_{c},\phi_{q})^T$ comprises the classical and quantum
Keldysh components of the boson modes describing excitations of the
quantum wires. The (inverse) Keldysh Green function $\hat
G^{-1}\equiv \left(
    \begin{smallmatrix}
      0& (G^-)^{-1}\cr
      (G^+)^{-1}& (G^{-1})^K
    \end{smallmatrix}
  \right)$ contains the advanced/retarded component
  $(G^\pm)_{q,\omega}^{-1} ={1 \over 2 \pi \nu} (-v q^2 - q (\omega\pm
  i\delta))$, where $\nu$ may be the filling fraction of an FQHE-bar,
  or a measure of the interaction strength of a quantum
  wire~\cite{FisherGlazman}, and the Keldysh component $ (G^{-1})^K =
  {i \delta \over \pi \nu} q F(q)$, where $F$ is related to the single
  particle distribution function; at equilibrium $F(q)=\coth(\hbar v
  q/2T)$, and $\delta(\omega-vq)F(q) \equiv
  F_\omega\delta(\omega-vq)$, where $F_\omega=\coth(\hbar
  \omega/2T)$. Hereafter we put $\hbar=1$, and set the electron charge
  $e_0=1$.

The charging action 
%
 \begin{equation}
S_C[\phi]\equiv {1\over 2C} \int dt\, \left( Q\sigma_1  Q\right),
\end{equation}
%
accounts for the finiteness of the electrostatic capacitance, $C$, of
the quantum dot. Here, $\sigma_1$ is a Pauli matrix, acting in Keldysh
space, and
%
\begin{equation}
 Q\equiv \sum_l  Q_l = {1\over 2\pi}( \phi_{3,d}+ \phi_{4,d}- \phi_{1,d}- \phi_{2,d}),
\end{equation}
%
is the total charge on the dot. It is given by the sum of the four charges $Q_l$
carried by the modes within the dot region, where $ Q_{1/2} =
-{1\over 2\pi}  \phi_{1/2,d},  Q_{3/4}={1\over 2\pi}
\phi_{3/4,d} $, $\phi_{ld} \equiv \phi_l(x_l) $ is the
field of the $l^{th}$ wire evaluated next to the lead-dot contact
point (cf. Fig. 1b), and a constant defining the charge
neutrality point has been ignored. 
 
The coherent scattering of quasi-particles from incoming leads,
$i=1,2$ to outgoing leads $o=3,4$ is described by the non-linear action
\begin{align}
\label{eq:S_tun}
&  S_{\mathrm{tun}}[\phi]=\sum_{i=1,2;o=3,4} \gamma_{ij} \sum_{s=\pm}\int dt \, \times \\
&  \times \cos\left(\phi_{i,d,c}(t) +  s {\phi_{i,d,q}(t)\over 2} - \phi_{o,d,c}(t) + s
  {\phi_{o,d,q}(t)\over 2}\right),\nonumber
\end{align}
where $\gamma_{ij}$ is the amplitude for elastic scattering from lead
$i$ to $j$ at 'scattering hotspots' within the dot\cite{Pnote}.
Finally, we assume that one of the incoming wires, $i=1$, is subject
to a bias voltage. We model the latter as a voltage kink of
time-modulated hight $V(t)$ and extension from $-L\to -\infty$ to
$-a<0$. The corresponding action reads $S_V[\phi_1] = \int
dt\,(\phi_{1,q}(-a,t)-\phi_{1,q}(-L,t)) V(t)$.  

Our next step is to integrate over the bosonic fields, barring the
QD-lead contact points, $x_l$.  We obtain a zero-dimensional action
\begin{equation}
\label{eq:S_reduced}
 S=S_\mathrm{dis}+S_{C}+S_{\mathrm{tun}}+\tilde S_V,
\end{equation}
where the dissipative action reads
\begin{align*}
   S_\mathrm{dis}[ \phi]={1 \over 2\pi \nu} \sum_l\int {d\omega\over
     2\pi} {i\omega\over 2\pi}
   \phi_{l,d,-\omega}^T
  \left(
    \begin{smallmatrix}
      &-1\cr 1 & 2F_\omega
    \end{smallmatrix}
  \right)  \phi_{l,d,\omega} \  ,
\end{align*}
and $\tilde S_V[\phi_1]=-2 \int {d\omega \over 2\pi} \phi_{1,q,-\omega}
V_{\omega}$.  In order to isolate the gapless modes of the problem, we transform to the basis
\begin{align}
\label{eq:transformation}
  \Phi_1 &= (\phi_{1,d}-\phi_{2,d}+\phi_{3,d}-\phi_{4,d})/2,\cr
\Phi_2 &= (\phi_{1,d}-\phi_{2,d}-\phi_{3,d}+\phi_{4,d})/2,\cr 
\Phi_3 &= (\phi_{1,d}+\phi_{2,d}-\phi_{3,d}-\phi_{4,d})/2,\cr 
\Phi_4 &= (\phi_{1,d}+\phi_{2,d}+\phi_{3,d}+\phi_{4,d})/2, 
\end{align}
where $ \Phi_1, \Phi_2 $ and $ \Phi_3$
appear as arguments of $S_{\mathrm{tun}}$ and
of $S_c$. Under renormalization they become
massive, hence irrelevant to the low energy dynamics. The latter is
dominated by the mode $\Phi\equiv \Phi_4$, whose action is 
%
\begin{equation}
    S[ \Phi]=2 \int {d\omega\over 2\pi}\,
    \left[{i\omega\over \pi \nu}
   \Phi_{-\omega}^T
  \left(
    \begin{smallmatrix}
   0    &-1\cr 1 & 2F_{\omega}
    \end{smallmatrix}
  \right) \Phi_{\omega}- \Phi_{q,-\omega}V_\omega\right].
  \label{massless.eq}
\end{equation}
%
The independency of this action on scattering parameters and dot
capacitance hints at universal behavior emerging in the low frequency
scaling limit. The origin of this universality is that $\Phi$ is a
soft mode related to overall charge
conservation in the system; unlike with $\Phi_{1,2,3}$ this mode is protected against scattering.

To define physical observables (e.g. current, response functions)
probing the universal physics of the system, we must refer to field
fluctuations at representative points (`observation points'), $\tilde
x_l$, $l=1,\dots,4$, on the leads (cf. Fig. \ref{fig:1}.) Expressed in terms of the native
fields, the current at these points is given by $j_{l,c,\omega}= {1
  \over 2 \pi} \partial_x \phi_{l,c,\omega}(\tilde x_l)$.  We aim to
compute current correlation functions in the low frequency limit where
only the universal mode $\Phi$ prevails. To this end, we employ the
identity
\begin{align}
\label{gaussian.eq}
 I_{l,\omega}&\equiv \langle j_{l,c,\omega}\rangle = \langle \Xi_{l,\omega}\rangle_{ \Phi},\\
  X_{ll',\omega}&\equiv \langle j_{l,c,\omega} j_{l',c,-\omega}\rangle= \langle
 \Xi_{l,\omega}\Xi_{l',-\omega}\rangle_{ \Phi},\qquad l\not=l',\nonumber\\
&\hspace*{-.4cm}\quad \Xi_{l,\omega}\equiv {1\over 2\pi} (\partial_{\tilde x_l} G(\tilde x_l-x_l) (G(0))^{-1} 
\Phi)_{c,\omega}+ {\nu\over 2 \pi} V_\omega\delta_{l,1},\nonumber
\end{align}
where we note that the coordinate representation of the lead Green
functions depends only on
coordinate differences and the angular brackets on the left and right
side denote functional averaging over the full action and the action
of the zero mode, respectively\cite{Cnote}. In the low frequency limit
$(x_1 - x_{1,d})\omega \to 0$, the Green functions do not depend on
positions explicitly anymore but still describe the causal relation
between different spatial points. The quadraticity of the Goldstone
mode action allows us to compute the correlation functions
(\ref{gaussian.eq}) explicitly.

\textit{Conductance ---}. Evaluating the first of the correlation
functions (\ref{gaussian.eq}) for the biased incoming lead, $l=i=1$, we
find $I_{1,\omega}= {\nu\over 2 \pi}
V_\omega $, while, evidently, there is no incoming current in the other lead,
$I_{2,\omega}=0$.  As for the outgoing currents ($o=3,4$) we
obtain a linear current voltage characteristic
\begin{equation}
\label{eq:outgoing_current}
I_{o,\omega}=  {\nu \over 2 \pi}  {V_\omega\over 2},
\end{equation}
which implies current conservation and conductance coefficients 
\begin{equation}
\label{eq:conductance}
{\cal G}_{io}={1 \over 2} { \nu \over 2 \pi}, 
\end{equation}
between incoming and outgoing leads.  In other words: the gapless mode
$ \Phi$ describes an $({1 \over 2},{1 \over 2})$-beam splitter. For
finite temperatures/AC-frequencies, the conductance coefficients will
show scale dependent corrections to the $({1 \over 2},{1 \over 2})$
limit, which non-universally depend on the bare
values of coupling constants\cite{Snote}.

\textit{Equilibrium noise ---.} 
We next analyze noise and cross-current correlations in the system. 
 In thermal equilibrium, {\it i.e.}  no external voltage bias, $V=0$, we
 obtain the intra-wire noise by employing a generalization of the identity 
 Eq.(\ref{gaussian.eq}) to the case of two fields in the same wire~\cite{Cnote} and find
\begin{align}
 \label{eq:thermal_intra}
X_{ll,\omega}= {\nu\over 2 \pi}
  \omega F_\omega
\end{align}
for all $l=1,\dots,4$.  Notice that the combination $\omega F_\omega$
crosses over from $|\omega|$ for $|\omega|>T$ to thermal scaling $2 T$
for $|\omega|<T$. We thus conclude that the low frequency intra wire
noise in our system is thermal.  For the cross-wire correlations we
use   Eq.(\ref{gaussian.eq}) and  obtain the following results:
\begin{align}
 \label{eq:thermal_inter}
  \mbox{outgoing/outgoing}&\; X_{oo',\omega}=0,\cr
  \mbox{incoming/outgoing}&\; X_{io,\omega}= {  \nu \over 4 \pi}\,
  \omega F_\omega,\cr
\mbox{incoming/incoming}&\; X_{ii',\omega}=0.
\end{align}
The intriguing observation here is the absence of correlations between
different outgoing and incoming wires. For the incoming wires, this
result appears intuitive: there is no causal connection between
different wires, i.e. two different incoming wires simply do not know
about each other. For the outgoing leads this is less
evident. However, outgoing and incoming wires can be mapped onto each
other by a time reversal operation, and this suggests that their
(equal time) fluctuation behavior should be identical.  We note that
Eqs. \eqref{eq:conductance}, \eqref{eq:thermal_intra}, and
\eqref{eq:thermal_inter} satisfy the fluctuation-dissipation
theorem. Finally, current conservation, $\sum_{i=1}^2 j_{i,c,\omega} =
\sum_{o=3}^{4} j_{o,c,\omega}$, requires that for any fixed index
$i_0=1,2$, $\sum_{i=1}^2 X_{i_0i}=\sum_{o=3}^{4} X_{i_00}$, a sum-rule
manifestly fulfilled by Eqs. (\ref{eq:thermal_intra}) and
(\ref{eq:thermal_inter}).

 \textit{Nonquilibrium noise ---}.  Here we are back to the situation
 where one of the incoming wires, $i$, is voltage biased.  
 We first
 discuss current correlations between the incoming wires. There are no
 correlations between 
 the biased incoming mode and
  the grounded mode.  Turning to correlations between the incoming biased and the outgoing
wires, we find
\begin{eqnarray}
\label{eq:4}
  X_{io,\omega}
 &=&   {\nu \over 4 \pi} \left({ \omega F_\omega } +  {\nu \over 2 \pi} |V_\omega|^2 \right),
\end{eqnarray} 
which fixes the incoming auto-correlation 
\begin{equation}
\label{eq:5}
X_{ii,\omega}= {\nu \over 2 \pi} \left(\omega F_\omega +{\nu \over 2 \pi}  |V_\omega|^2  \  \delta_{i,1} \right).
\end{equation}
by current conservation. Eqs. (\ref{eq:4},\ref{eq:5}), and our
above results on the average current determine the cumulants, $S_{ll'} \equiv X_{ll'} -
\langle I_l\rangle \langle I_{l'}\rangle$ as
\begin{align}
  \label{eq:16}
  S_{io,\omega} =  {\nu \over 2 \pi} {\omega F_\omega\over 2},\qquad
  S_{ii,\omega} =  {\nu \over 2 \pi} {\omega F_\omega }.
\end{align}
Remarkably, this result coincides with Eq.\eqref{eq:thermal_inter}. 
In other words: the noise is purely thermal, there is no shot noise in
the incoming wire, and no $V$-dependent correlations to the current in
the outgoing wires~\cite{foot_T}.

Turn to correlations in the outgoing
  wires, we apply Eq. (\ref{gaussian.eq}) once more to obtain
\begin{eqnarray}
  X_{oo',\omega} 
  &  =  &\left( {\nu \over 4 \pi} \right)^2   |V_\omega|^2 \nonumber\\ 
  X_{oo,\omega} &=& \left( {\nu \over 4 \pi} \right)^2 |V_\omega|^2 +  { \nu \over 2 \pi}{\omega F_\omega}
\end{eqnarray}
for the inter- ($o\not=o'$) and intra- ($o=o'$) wire correlations, resp.  As
in the bias-neutral case above, these results respect current conservation.
Subtraction of the average currents yields the noise cumulants ($o\not=o'$)
\begin{align}
  \label{eq:17}
  S_{oo,\omega}= {\nu \over 2 \pi} {\omega F_\omega},\qquad
  S_{oo',\omega}= 0,
\end{align}
\textit{i.e.} in spite of the splitting of the incoming current into
two outgoing channels, the corresponding noise remains thermal, and
there are no inter-wire correlations. 

The picture above is based on a rather robust physical mechanism: of
the four modes supported by the two incoming and two outgoing wires,
three get frozen by a combination of quasiparticle scattering and
interaction, or, more technically, the simultaneous presence of more
than two independent and relevant contributions to the effective
action of the dot. While the  freezing of all relative fluctuations, $\Phi_{1,2,3}$, in the
system is
responsible for the $({1\over 2},{1\over 2})$ division of conductance
coefficients, one collective mode, $\Phi$, is
protected by current conservation. (In fact, $\Phi$ may be interpreted as
the Goldstone mode corresponding to the gauge fixing of the boson
fields.)  
The quadratic nature of the  action $S[\Phi]$, 
Eq. (\ref{massless.eq}), signals the absence of 'charge quantization'
effects in the low frequency dynamics of the system. In particular, it
implies the absence of noise, beyond the
'thermal noise' carried by the distribution $F$. However, this 
result requires careful consideration: at finite $V$, the quantum
scatterer, is kept in a non-equilibrium steady state, implying that
$F$ might be characterized by an effective, voltage dependent,
temperature, $T_\mathrm{eff}(V)$. The
absence of shot noise notwithstanding, this $T_{\mathrm{eff}}(V)$ may give rise to
voltage dependent current fluctuations, similar to those caused by
shot noise. However, genuine shot noise would also generate non-vanishing  cross-current correlations in the outgoing channels
\cite{anti}. The vanishing of these, therefore, has smoking gun status
in signaling the absence of shot noise in our system.

The generality of this picture suggests various candidates to confirm
its predictions in experiment. One example would be a quantum Hall
strip (in the fractional regime) a finite section of which has been tuned
to be in a compressible filling factor by gates. Another
intriguing possibility is that the value of $1/2$ observed for the conductance
of inhomogeneous graphene p-n junctions \cite{snake} is due to the
survival of only a single transmitting mode, along the lines of the
mechanism discussed here.

We acknowledge useful discussions with P. Brouwer, B. Halperin, and M. Heiblum.
This work was supported by BMBF, GIF,
 ISF, Minerva Foundation, SFB/TR 12 of the Deutsche Forschungsgemeinschaft, and  EU GEOMDISS.





\begin{thebibliography}{99}
\bibitem{Pnote} The detailed position of
these points does not qualitatively affect our results, whence we
identify them with the points $x_l$ above.
\bibitem{Cnote}
To prove these identities, one has to carry out the
Gaussian integration leading to (\ref{eq:S_reduced}) in the presence
of the pre-exponential term on the l.h.s. As a result of a
straightforward if somewhat lengthy calculation one obtains the
r.h.s. The current auto-correlation function, $l=l'$ is given by a
slightly more complicated expression which we do not state explicitly,
as its evaluation is fixed by current conservation.
\bibitem{foot_T} This result tacitly assumes that the whole system is
  kept at a thermal distribution characterized by some effective
  temperature $T$. Heating effects may influence this temperature.
\end{thebibliography}
\end{document}